\documentclass[fleqn,usenatbib]{mnras}

\usepackage[T1]{fontenc}

\DeclareRobustCommand{\VAN}[3]{#2}
\let\VANthebibliography\thebibliography
\def\thebibliography{\DeclareRobustCommand{\VAN}[3]{##3}\VANthebibliography}

\usepackage{graphicx}	
\usepackage{amsmath}	
\usepackage{amssymb}	

\usepackage{algorithm}
\usepackage[noend]{algpseudocode}

\usepackage{newtxtext,newtxmath}

\DeclareMathOperator*{\argmin}{arg\,min}

\title[Lucky Imaging without the Waste]{\textit{The Thresher} : Lucky Imaging without the Waste}

\author[J. A. Hitchcock et al.]{
J. A. Hitchcock $^{1}$\thanks{E-mail: jah36@st-andrews.ac.uk (JAH)},
D.M. Bramich $^{2,3}$,
D. Foreman-Mackey $^{4}$,
David W. Hogg $^{4}$
and M. Hundertmark $^{5}$
\\
$^{1}$ SUPA, School of Physics and Astronomy, University of St. Andrews, North Haugh, KY16 9SS, UK\\
$^{2}$ Center for Astro, Particle and Planetary Physics, New York University Abu Dhabi, P.O. Box 129188, Saadiyat Island, Abu Dhabi, UAE \\
$^{3}$ Division of Engineering, New York University Abu Dhabi, P.O. Box 129188, Saadiyat Island, Abu Dhabi, UAE \\
$^{4}$ Center for Computational Astrophysics, Flatiron Institute, New York, NY \\
$^{5}$ Astronomisches Rechen-Institut, Zentrum f{\"u}r Astronomie der Universit{\"a}t Heidelberg (ZAH), 69120 Heidelberg, Germany 
}

\date{Accepted XXX. Received YYY; in original form ZZZ}

\pubyear{2015}

\begin{document}
\label{firstpage}
\pagerange{\pageref{firstpage}--\pageref{lastpage}}
\maketitle

\begin{abstract}
 In traditional lucky imaging (TLI), many consecutive images of the same scene are taken with a high frame-rate camera, and all but the sharpest images are discarded before constructing the final shift-and-add image.  Here we present an alternative image analysis pipeline ---\textit{The Thresher}--- for these kinds of data, based on online multi-frame blind deconvolution. It makes use of all available data to obtain a best estimate of the astronomical scene in the context of reasonable computational limits; it does not require prior estimates of the point-spread functions in the images, or knowledge of point sources in the scene that could provide such estimates. Most importantly, the scene it aims to return is the optimum of a justified scalar objective based on the likelihood function. Because it uses the full set of images in the stack, \textit{The Thresher} outperforms TLI in signal-to-noise; as it accounts for the individual-frame PSFs, it does this without loss of angular resolution.  We demonstrate the effectiveness of our algorithm on both simulated data and real Electron-Multiplying CCD images obtained at the Danish 1.54m telescope (hosted by ESO, La Silla). We also explore the current limitations of the algorithm, and find that for the choice of image model presented here, non-linearities in flux are introduced into the returned scene. Ongoing development of the software can be viewed at \url{https://github.com/jah1994/TheThresher}.

\end{abstract}




\section{Introduction}

The atmosphere produces high-frequency temporal and spatial variations in the point-spread function (PSF) of ground-based astronomical images, as turbulent convective cells pass over the telescope aperture. The time-averaging effect of fluctuations of the PSF for conventional observations quickly degrades the resolution, and results in blurry, band-limited images. A class of techniques for compensating for this involve obtaining a large number of extremely short exposures close to the timescale of the atmospheric variations. These short exposures individually have low signal-to-noise and very complicated PSFs, but they provide strong constraints on the diffraction limited scene when many images are used collectively.

The idea of exploiting the high resolution information in short
exposures was first suggested by \citet{labeyrie1970attainment} and there has been a rich literature on this topic since. In particular, a very popular technique which we will call \emph{traditional} lucky imaging (TLI) \citep{law2006lucky} is based on Fried's derivation of the probability of obtaining a frame with unusually ``lucky'' seeing\footnote{A measure of the blurring of an astronomical image due to atmospheric turbulence.} when many images are taken quickly. That is, occasionally---just by chance---the wavefront distortions from the atmosphere will come close to cancelling the imperfections in your telescope.  These best images are identified, shifted, and co-added, with the rest of the data relegated to the dustbin. TLI is very popular because it is simple to understand and implement, it is computationally tractable, and it can be used with very inexpensive equipment.  As a technique, however, it is not \emph{really} inexpensive because it results in a substantial amount of wasted data. Typical TLI results are based on only the best (i.e. sharpest) percentage of the acquired images, and progress has focused on improving the way images are co-added or selected  to improve the efficiency of the method \citep[for example]{staley2010data, mackay2013high}.

The work presented here was motivated by the (correct) feeling that, treated properly, there is no way that the discarded majority of data in a TLI imaging stack can be detrimental to constraining the astronomical scene!  In the context of TLI, the fundamental reason why throwing away data \emph{helps} is that the core data analysis step is shift-and-add co-addition of the imaging, which (though widely used in astronomy) is not justifiable when the point-spread function is varying rapidly.

The method we present here---\textit{The Thresher}---is a new flavour of an old idea.  It is a special case of a broad class of algorithms called \emph{blind deconvolution} \citep[]{ayers1988iterative, campisi2017blind}. It is closely based on the online multi-frame blind deconvolution (OMFBD) image analysis method of \citet{hirsch2011online} in particular. However, our algorithm differs from this prior work in several important ways: 1) \textit{The Thresher} implements a physically motivated, justifiable likelihood function for short-exposure images; 2) our algorithm implements a robust stochastic gradient descent procedure based on state-of-the-art optimisation algorithms, and 3) It is entirely general to the choice of image model and likelihood function, as it makes use of automatic differentiation tools. Points 1) and 2) in particular make \textit{The Thresher} extremely well suited to dealing with realistically faint and noisy high frame-rate astronomical imaging data. 

In this work, we describe the theory behind our algorithm and the details of our particular implementation. Most importantly, we demonstrate the significant improvements over TLI made possible by this technique using both simulated and real data from Electron-Multiplying (EM) CCD, fast imaging cameras. A fundamental limitation of TLI is that the signal-to-noise of the final co-add is \emph{inversely related} to its resolution (i.e. the better the resolution, the poorer the signal-to-noise). We show that this need not be the case; \textit{The Thresher} has the potential of returning an image with the signal-to-noise of the entire data set at the resolution of the very best images.

This manuscript is structured as follows. In Section \ref{sec:problem_formulation}, we describe the inference problem and the algorithm developed to solve it. We present our results on simulated and real data in Sections \ref{sec:simulated} and \ref{sec:real}, respectively. In Section \ref{sec:limits}, we probe current limitations of the algorithm and outline the scope for further improvements. Section \ref{sec:conclusion} states our conclusions.

Finally, in order to help the reader keep track of the notation introduced in the following sections, we include a table of symbols and their definitions in the Appendix (Table \ref{tab:notation}).

\section{Problem Formulation}\label{sec:problem_formulation}

\subsection{Online Multi-frame Blind Deconvolution}\label{sec:mobd}

We start by writing down a model for our imaging data. For any short exposure, $y_n$, in our Lucky Imaging (LI) data set of $N$ images, a reasonable model for the light distribution at any given $ij$ pixel is
\begin{equation}
    m_{n, ij} = [k_n \otimes s]_{ij} + b_n \;.
    \label{eq:modelimage}
\end{equation}
Equation \ref{eq:modelimage} states that each image is generated by the convolution of some high-resolution, `true' image $s$, with some unknown blur kernel, $k_n$, and some additive sky background, $b_n$. Throughout this work, we assume that $s$ is time invariant, and that both $k_n$ and $b_n$ are spatially invariant. The latter is justified for the small angular area LI data that we use -- on the order of the scale of the isoplanatic patch -- where the seeing is approximately constant over the field-of-view.

Given some statistical model for the \textit{noise} in our imaging data, we can write down a likelihood function for $y_n$ given our vector of model parameters, $\boldsymbol{\theta} = [k_1, ... , k_N, b_1, ... , b_N, s]$. The notation is kept light for clarity. For each $y_n$ image, $k_n$ is modelled as a square array of pixels of some user-specified size, and $b_n$ is a scalar. $s$ is modelled as an array of pixels with the same dimensions as the data images. Assuming independence between images, and independence between pixels, the likelihood function for our entire data set, $\boldsymbol{y} = [y_1, ..., y_N]$, is just the product of the individual pixel likelihoods,
\begin{equation}
    p(\boldsymbol{y}|\boldsymbol{\theta}, \boldsymbol{\Omega}) = \prod_{n=1}^{N} \prod_{ij} p(y_{n, ij}|k_n, b_n, s, \boldsymbol{\Omega}) \;.
    \label{eq:prod_likelihood}
\end{equation}
We introduce $\boldsymbol{\Omega}$ to represent the list of other known quantities the likelihood is conditional on e.g. the parameters of the detector noise model.

For numerical convenience, it is helpful to instead work with logarithms, which transforms the above product into a sum. We can then define the total \textit{loss} as the negative log-likelihood of our imaging data given the model parameters
\begin{equation}
\begin{aligned}
    L(\boldsymbol{y}; \boldsymbol{\theta}, \boldsymbol{\Omega}) &=  - \ln p(\boldsymbol{y}|\boldsymbol{\theta}, \boldsymbol{\Omega}) \\
    &= - \sum_{n=1}^{N} \sum_{ij} \ln p(y_{n,ij} | k_n, b_n, s, \boldsymbol{\Omega}) \\
    &= \sum_{n=1}^{N} l(y_n ; k_n, b_n, s, \boldsymbol{\Omega}) \;,
\end{aligned}
\label{eq:total_loss}
\end{equation}
with $l(y_n ; k_n, b_n, s, \boldsymbol{\Omega})$ being the loss associated with a single image.

Unfortunately, there are a couple of practical complications that prevent us from straightforwardly minimising Equation \ref{eq:total_loss} to obtain the maximum-likelihood estimate (MLE) for our model parameters. Firstly, LI data sets consist of thousands of images, and it is not practical to load these all into computer memory simultaneously. Secondly, and most importantly, in this blind deconvolution setting, for each $y_n$ image, $k_n$ and $s$ are perfectly degenerate, as their exists a
limitless number of possible combinations of these parameters that
could generate the observed data (Equation \ref{eq:modelimage}).

\citet{hirsch2011online} proposed an iterative \textit{online} algorithm, which needs to access only a single image at any time, and makes the problem tractable, consisting of two steps for each $y_n$ image in our data set:
\begin{equation}
    \mathrm{(i)} \; [\hat{k}_n, \; \hat{b}_n] = \argmin_{k_n \geq 0, \; b_n} l(y_n ; k_n, b_n, s_t, \boldsymbol{\Omega})
    \label{eq:step1}
\end{equation}
\begin{equation}
    \mathrm{(ii)} \; s_{t+1} \gets  s_t - \alpha_t \nabla_{s_t} l(y_n ; \hat{k}_n, \hat{b}_n, s_t, \boldsymbol{\Omega}) \;.
    \label{eq:step2}
\end{equation}
In step (i), the current estimate of the scene, $s_t$, is kept fixed, and the unique $k_n$ and $b_n$ which minimise the loss for the given $y_n$ are inferred. In step (ii), these new estimates for the blur kernel and sky background are used to re-evaluate the loss. The gradient of the loss is computed with respect to $s_t$, which is then updated with a single steepest descent step with step-size $\alpha_t$, to give us a new estimate for the high-resolution scene, $s_{t+1}$. In this way, we can sequentially update our estimate of $s$ by repeating steps (i) and (ii) for every image we can access. If required, multiple $I$ passes can be made over the imaging data (i.e. giving a total of $I \times N$ updates).

As $k_n$ and $b_n$ are unique to each $y_n$, these should be inferred by minimising the loss for any given image. $s$ however is common to all images, and so only a single update should be performed in step (ii). This update is an example of stochastic gradient descent (SGD; see \citealt{bottou2018optimization} for an excellent overview). For settings in which it is impractical to compute the gradient of the total loss with respect to our model parameters, $\nabla L$, SGD allows us to make progress by computing \emph{stochastic} approximations to the total gradient, $\nabla l$. Despite being noisy approximations to $\nabla L$, optimisation by SGD typically makes rapid initial progress, as when far from the optimum, $\nabla l$ will very likely have the same sign as $\nabla L$. This progress will generally be very sensitive to the step-size, particularly once in the vicinity of the optimum, but with some appropriately decaying step-size\footnote{This is hugely problem specific, and typically requires empirical tuning. Despite SGDs popularity and long history, the optimal choice of step-size for a given optimisation problem is an open research problem.}, SGD can be shown to converge in both convex and non-convex settings \citep{bottou1998online}. In practice however, it is useful to terminate the optimisation early, as issues with model representation can occur as the scene is deconvolved.

In order to stabilise and automatically anneal the stochastic updates to $s$, we propose a practical modification to step (ii) by introducing the following update into our SGD step,

\begin{equation}
    s_{t+1}\gets  s_t - \frac{\alpha_0}{\sqrt{\hat{\nu}_t} + \epsilon} \hat{\mu}_t \;,
    \label{eq:Adam_update}
\end{equation}

where $\hat{\mu}_t$ and $\hat{\nu}_t$ are the (unbiased) exponentially decaying running averages of the gradient and squared gradients respectively, and serve as estimates of the mean and (uncentered) variance of the gradients. This is an example of the popular Adam update, which adaptively tunes the learning rate for the parameters in a way that is sensitive to their gradient history.

We refer the reader to \citet{kingma2014adam} for definitions of how $\hat{\mu}_t$ and $\hat{\nu}_t$ are computed. Therein, we adopt the suggested values for the hyper parameters. This includes the value for $\epsilon$, which is a small number added to the denominator of Equation \ref{eq:Adam_update} to improve numerical stability.

It is useful to think of the quantity $\hat{\mu}_t / \sqrt{\hat{\nu}_t}$ as something like the signal-to-noise ratio (SNR) of the gradients. The Adam update builds \emph{momentum} for parameters with persistent non-zero gradients, and exerts friction when the variance of past gradients grows. This is useful, since parameters with a low `SNR' should be characterised by oscillating about some optimum (or they are just insensitive to the data), and so their step-size decreases automatically.

For our problem, $s$ has many free parameters -- modelled as pixels on a grid -- associated with both the sky background and astronomical sources of interest. Provided we can reliably estimate the sky level with our model (Equation \ref{eq:modelimage}), the gradient histories of pixels associated with only this smooth background (i.e. regions in $s$ not populated with sources) should typically be centered around 0. Consequently, their effective step-sizes are naturally annealed by the update in Equation \ref{eq:Adam_update}. Conversely, pixels associated with sources should have a greater consistency in gradient sign and lower variances, and so their effective step-size is annealed less aggressively. This has the overall effect of stabilising the deconvolution, as it helps to suppress noise amplification\footnote{Inaccuracies in either step (i) or (ii) will propagate to the next step in a positive feedback loop.} in the updates to $s$.

Another major advantage of this SGD framework is that the only hyperparameter that needs some empirical tuning is the learning rate, $\alpha_0$. This can be set independently for every parameter, and for this problem, we strongly recommend that it is! As argued above, the pixels associated with astronomical sources in $s$ should be allowed to vary the most -- for this is where the data are most informative -- and so these parameters should be assigned larger initial learning rates. Further, $\alpha_0$ has the added advantage of approximately bounding the size of the per-parameter updates for each SGD step. In this way, it performs a function very similar to the update clipping scheme used by \citet{lee2017robust} to stabilise the deconvolution of noisy astronomical images.

\subsection{A Poisson-Gamma-Normal Noise Model for EMCCD Data}

Having written down a model for our imaging data, and an algorithm for inferring its parameters, we now need to choose a suitable loss function (i.e. a negative log-likelihood) that represents our belief about the observation noise. In this work, we consider observations acquired with Electron-Multiplying CCDs (EMCCDs).

A distinguishing feature of astronomical scenes is their huge dynamic range; some objects are very bright, and many more are typically very faint. Objects of interest in individual LI images will exist at very low signal-to-noise, with many below a reasonable detection threshold in any given exposure. Furthermore, collections of sufficiently short LI exposures are richly informative, containing high-frequency (spatial) information otherwise lost in conventional, band-limited, long exposures. These high-frequency components in general exist at the lowest intensities, and will therefore be very sensitive to the noise. For all but the very brightest sources, in these photon-starved images, Gaussian approximations to Poissonian counting statistics do not apply, and as will be seen, the electron-multiplication process particular to EMCCDs introduces an additional source of noise. An accurate statistical model which incorporates our \emph{physical} knowledge of the noise generating processes of EMCCDs should allow us to better exploit this information, and take full advantage of the maximum-likelihood method for making robust and accurate inferences from our imaging data. 
 
\citet{korevaar2011maximum} and \citet{hirsch2013stochastic} independently derived the same physically motivated likelihood function for EMCCD data. This probability density function (PDF) is built from convolutions of 1) Poissonian photon noise and spurious charge events, 2) The cascade amplification of charge in the EM register, which is well approximated by a Gamma distribution, and 3) Normally distributed readout noise. After setting $\boldsymbol{\Omega} = [f, G, \sigma_0, c, q]$ (refer to Table \ref{tab:detector_params} for definitions of these detector parameters), in units of ADU$^{-1}$, this Poisson-Gamma-Normal (PGN) likelihood takes the form,
\begin{multline}
    p_{\mathrm{PGN}}(y_{n, ij}|k_n, b_n, s, \boldsymbol{\Omega}) = \\
    f\,H[y_{n,ij}]\,
    \sqrt{\frac{\lambda_{n, ij}}{y_{n, ij}\, G^2}}\,\,I_{1}\left[\frac{2f}{G}\sqrt{\lambda_{n, ij}\,y_{n,ij}}\right]\,\,\exp \left[-\frac{f}{G}(\lambda_{n,ij} + y_{n, ij})\right] +  \\ \frac{f}{\sqrt{2\pi}\,\sigma_0}\,\,\text{exp}\left[-\frac{1}{2}\left(\frac{f\,y_{n,ij}}{\sigma_0}\right)^2\right]\,\,\text{exp}\left[-\frac{f}{G}\lambda_{n,ij}\right]\;,
    \label{eq:PGN_likelihood}
\end{multline}
 where our image model (Equation \ref{eq:modelimage}), enters through
\begin{equation}
    \lambda_{n, ij} = q m_{n, ij} + (G/f)c \;.
    \label{eq:lambda}
\end{equation}

Note that $H$ is the Heaviside step function and $I_1$ is the modified Bessel function of the first order.

For individual $y_n$ exposures and our image model (Equation \ref{eq:modelimage}), the loss function under this noise model is simply the sum of the per-pixel negative log-likelihoods,
\begin{equation}
    l_{\mathrm{PGN}}(y_{n} ; k_n, b_n, s, \boldsymbol{\Omega}) =
    - \sum_{ij} \ln p_{\mathrm{PGN}}(y_{n, ij} | k_n, b_n, s , \boldsymbol{\Omega}) \;,
    \label{eq:PGNloss}
\end{equation}
and the total loss over all images is equal to
\begin{equation}
    L_{\mathrm{PGN}}(\boldsymbol{y} ;\boldsymbol{\theta}, \boldsymbol{\Omega}) = \sum_{n=1}^{N} l_{\mathrm{PGN}}(y_{n} ; k_n, b_n, s, \boldsymbol{\Omega}) \; .
\end{equation}

\begin{table}
    \centering
    \begin{tabular}{c|c}
    \hline
    Parameter & Description and units  \\
    \hline
    f & A/D conversion factor ($e_{\mathrm{EM}}^{-}$ / ADU) \\
    G & Electron-multiplying (EM) gain ($e_{\mathrm{EM}}^{-}$ / $e_{\mathrm{Photon}}^{-}$) \\
    $\sigma_0$ & Readout noise ($e_{\mathrm{EM}}^{-}$) \\
    c & Spurious charge ($e_{\mathrm{Photon}}^{-}$) \\
    q & Quantum efficiency (dimensionless) \\
    \end{tabular}
    \caption{EMCCD detector parameters. We adopt the same notation as \citet{harpsoe2012high}, who differentiate between electrons generated before and after the EM amplification as $e_{\mathrm{Photon}}^{-}$ and $e_{\mathrm{EM}}^{-}$ respectively.}
    \label{tab:detector_params}
\end{table}

\subsection{Noise Model Validation and Calibration}\label{sec:noisemodelval}

We will now assess how well the PGN noise model can capture the noise properties of real EMCCD images acquired with the Danish 1.54m (DK154) TCI `red' camera \citep{skottfelt2015two}. Of course, real world data brings real world problems, and the distribution of DK154 image counts on \emph{raw} images stored for analysis is not well represented by Equation \ref{eq:PGN_likelihood}.

Firstly, due to storage constraints, despite being readout at 16-bit precision, the pixel values in the raw images are quantised to integers. We therefore `dither' the raw images prior to further calibration by adding to each integer pixel value a float randomly sampled from the uniform distribution, $\mathcal{U} \sim (-0.5, +0.5)$. This replaces the discretised raw image with a plausible analogue representation as originally observed by the camera. Under the values of $f$ and $G$ that apply to DK154 data, the associated uncertainty introduced by this procedure is approximately an order magnitude smaller than the readout noise, and can safely be ignored.

Secondly, as with conventional CCD images, systematic bias and inter-pixel quantum efficiency differences should be corrected for. After dithering, we therefore bias subtract and flat-field all EMCCD images. However, one additional step must be taken to calibrate the EMCCD images, as there is a significant bias level drift -- caused by on-chip heat-dissipation associated with the fast readout and large current -- unique to each and every image, that should be corrected. This drifting bias level can be measured by comparing the values of the pixels in the unilluminated overscan regions of the given image and the master bias frame\footnote{Every $(i,j)$ pixel value in the master bias frame is the truncated mean of the corresponding pixel values from 1000 bias images, where the top 5\% of counts for the given pixel are rejected. The master bias frame can be considered noiseless, since the mean pixel variation between master bias images is an order of magnitude lower than the readout noise (Section 5.2.1 of \citet{skottfelt2015two})}. As these pixels too are affected by the cascade EM amplification, \citet{harpsoe2012high} proposed using the truncated mean of the distribution of the difference in these pixel counts as an estimate of the bias level drift. Specifically, the truncated mean is taken to be the mean computed after rejecting the top 5\% of counts, and this value is subtracted from the relevant science frame. 

We note that, in principle, one could introduce another free parameter into our detector model to represent this bias level drift, whose value could then be optimised, which would leverage the power of our physically motivated noise model. This is definitely more accurate than the heuristic approach outlined above, but it is challenging to implement in practice as the likelihood surface with respect to this parameter is discontinuous -- since the likelihood is piece-wise about a point determined by this parameter -- and so non-gradient based optimisation approaches would need to be used.

\subsubsection{Calibrating the noise model parameters}\label{sec:calibration}

We took a sequence of 500 DK154 dark images with the same camera settings that are used throughout this work. From these we can calibrate all but one of our noise model parameters. As these images are not illuminated, any `signal' can be assumed to be solely due to spurious charge events, which is equivalent to substituting $\lambda_{n, ij} = (G/f)c$ into Equation \ref{eq:PGN_likelihood}. Consequently, we cannot determine $q$ -- which is in general very challenging to measure -- and so we adopt the detector manufacturer's value of $q = 0.8$ throughout this work.

\begin{figure}
    \centering
    \includegraphics[width=\linewidth]{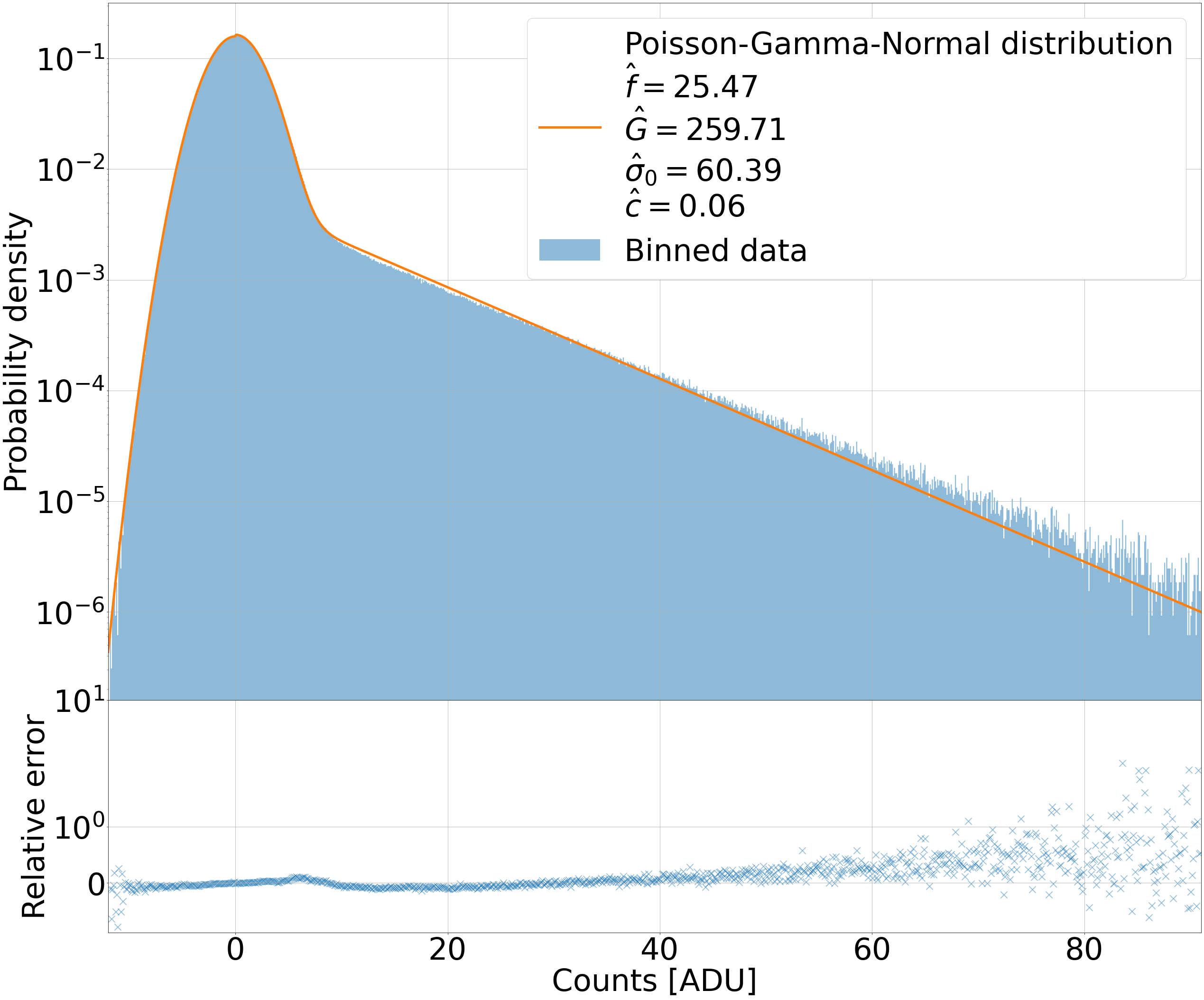}
    \caption{Normalised histogram of pixel counts from 500, $256\times 256$ pixel (i.e. $\sim3.3\times 10^7$ data points in total), dark DK154 EMCCD images, overlain with the fitted Poisson-Gamma-Normal (PGN) likelihood. The positive fat tail is a result of amplified spurious charges. The relative difference between the model and data -- computed as (data - model) / model -- for each bin is plotted in the lower panel. The MLE for the detector parameters are also shown (see Table \ref{tab:detector_params} for units).}
    \label{fig:PGN_validation}
\end{figure}

For a stack of $n=1, ..., 500$ dark images acquired with an EMCCD camera with detector parameters $\boldsymbol{\Omega} = [f,G,\sigma_{0},c,q]$, we minimise the following objective function, 
\begin{equation}
\begin{aligned}
    \left[\hat{f}, \hat{G}, \hat{\sigma}_0, \hat{c}\right] &= \argmin_{f, G, \sigma_0, c} L_{\mathrm{PGN}}(\boldsymbol{y} ; \boldsymbol{\Omega})\\
    &= \argmin_{f, G, \sigma_0, c} \sum_{n=1}^{N} l_{\mathrm{PGN}}(y_{n} ; \boldsymbol{\Omega}) \;, \\
    &= \argmin_{f, G, \sigma_0, c} - \sum_{n=1}^{N} \sum_{ij} \ln p_{\mathrm{PGN}}(y_{n, ij} | \boldsymbol{\Omega}) \;,
\end{aligned}
\label{eq:validation_loss}
\end{equation}
where all free detector parameters are modelled as unknown constants. As there is a single readout register, this is a reasonable approximation for $f$, $G$ and $\sigma_0$. For the purposes of this work, we restrict our analysis to a $256\times256$ pixel central part of the DK154 camera, over which $c$ is also approximately constant (see Figure 3 in \citet{harpsoe2012high}). Equation \ref{eq:validation_loss} was optimised with the Adam algorithm (which is outlined in Section \ref{sec:mobd}, but here we are using the full gradient, $\nabla L_{\textrm{PGN}}$, when updating). The model PDF (Equation \ref{eq:PGN_likelihood}) parameterised with the MLEs is overlain onto a normalised histogram of the data in Figure \ref{fig:PGN_validation}, with bin widths determined using the approach in \citet{knuth2006optimal}. The range of data to be binned was restricted to that with a predicted probability density of at least $10^{-6}$ under the MLE of the noise model. As there are $\sim 3.3\times10^7$ data points in total, this guards against large numbers of empty (or near-empty) bins. The relative difference between the model and binned data is plotted in the lower panel. We can see in Figure \ref{fig:PGN_validation} that the PGN noise model captures the asymmetric, and heavy-tailed features of our data distribution arising from the EM amplified spurious charges. However, we stress that in no way should this model be thought of as the `truth'; it is only an approximation. It is however a practically effective, physically motivated, and broadly accurate representation of our data. 

Now equipped with an accurate, detailed noise model for the detector -- with calibrated instrument parameters -- we can proceed to tackle the deconvolution problem.

\subsection{Constraints and Regularisation}

Due to the ill-posed nature of the blind deconvolution problem, it can be useful to include constraints and regularisation/penalty terms (i.e. Bayesian priors) on the model parameters to restrict the scope of their possible solutions, or push the model parameters to some \textit{a priori} preferred value when the data are uninformative.

\subsubsection{Non-negativity}

In the deconvolution literature, it is very common to restrict the pixel values in the underlying scene $s$ to non-negative values. For our problem, while it is reasonable to restrict the pixels in the blur kernel to non-negative values, it is less clear whether it is correct to enforce non-negativity on $s$. In any given LI image, there will almost certainly exist faint sources at or below the detection threshold that blend into the sky background. In the units of the final inferred scene, it could be that these sources have $\textit{negative}$ flux, and will be strongly affected by any non-negativity constraint.

In practice, this decision should, at least partly, be motivated by the adopted noise model. It seems reasonable to force $s$ to be non-negative if using the PGN noise model, as this will guard against possible numerical issues with likelihood evaluations, since under this particular noise model, the image model $m_n$, should be non-negative everywhere; the PGN noise model does not permit negative photoelectrons\footnote{For this reason, we also restrict $b_n \geq 0$ when using the PGN noise model, especially as we always sky-subtract $s_0$ to mitigate the anti-correlation between $k_n$ and $b_n$. The differential background level between this sky-subtracted model and the data will be non-negative, justifying this constraint, which in combination with the non-negativity constraints on $k_n$ and $s$ will ensure $m_n$ is always non-negative.}. However, if using a simple squared error loss (as we do for tests on simulated, noiseless images in Section \ref{sec:noiseless_tests}), no such `physical' argument for non-negativity of $s$ can be made. In this case, the crucial point is that the data do not constrain the \emph{absolute} value of any given parameter in $s$, but only the \emph{relative} values of the parameters. Consequently, in this work, we \emph{do not} enforce non-negativity on $s$ in the tests on noiseless data (Section \ref{sec:noiseless_tests}), but we \emph{do} impose this constraint for the tests on data modelled with the PGN noise model (Sections \ref{sec:noisy_tests} and \ref{sec:real}).

\subsubsection{Penalised Maximum-Likelihood-Estimation}

Noise amplification is the perennial problem for many deconvolution algorithms, and this is only exacerbated by the extremely low signal-to-noise of our short exposure data. It is then common to include penalty terms in the loss function to guard against excessively noisy blur kernels and scene estimates.

Given the degenerate nature of the inference problem, the accuracy with which we can determine each of the $k_n$ kernels will strongly determine how we update $s$ at each step. Indeed, extensive empirical experience gained from developing and testing \textit{The Thresher} has shown us that almost all problems in the reconstruction of $s$ can be traced back to poor kernel inference. The delta-function basis used to model the kernel is highly flexible, but this comes at the cost of increased overfitting, resulting in excessively noisy kernels, particularly when the signal-to-noise is low \citep[provide some instructive examples from the highly analogous Difference Image Analysis literature]{becker2012regularization, bramich2016difference}. This is a problem, as noise in the kernel will propagate to successive estimates for the scene. To guard against this, we include the L1 norm on the kernel pixels as a penalty term to our loss function, which pushes kernel pixels towards 0 unless the data are informative. Consequently, we can re-write step (i) (Equation \ref{eq:step1}) as
\begin{equation}
    \mathrm{(i)} \; [\hat{k}_n, \; \hat{b}_n] = \argmin_{k_n \geq 0, \; b_n}  l(y_n ; k_n, b_n, s_t, \boldsymbol{\Omega}) + n_{\textrm{pix}} \, \phi || k_n ||_1 \;,
    \label{eq:step1_penalised}
\end{equation}
where $n_{\textrm{pix}}$ is the number of pixels in each $y_n$ image, and $\phi$ is a tuning constant which sets the strength of the regularisation, and must be set empirically. In practice, we find values in the range of $\phi = 0.001 - 0.1$ to be reasonable, with lower signal-to-noise data requiring stronger kernel regularisation.

It is common to also include penalty terms on the scene which reflect our beliefs about the properties of images. These penalty terms typically disfavour high-frequencies (e.g. sharp gradients) in the scene, and so have a smoothing effect, and can help suppress noise. Popular choices include the L1 or L2 norm on the image gradients. Unfortunately, these penalty terms actually \textit{favour} blurry scenes \citep{levin2011understanding}. Additionally, they are prone to introducing correlated artefacts into the image model, and their gradient fields produce complicated PSFs which pose challenges for the purposes of any further measurements on the reconstructed image. These issues pose serious challenges for the astronomer, and distinguishes our use-case from the general image reconstruction setting. In astronomy, the target of the deconvolution must almost always be useful for the purpose of making \textit{measurements}, and the PSF is typically one of the things that we are most interested in measuring! Having a practically useful model for the PSF is extremely important for a variety of downstream tasks that we are likely to care about (e.g. source detection, astrometry, flux measurements).

Lastly, as discussed in \citet{vio2005least}, the penalisation of the high-frequencies of the image model could effectively nullify our efforts in implementing an accurate statistical description of the noise which attempts to exploit this low intensity, high-frequency information. These penalty terms are designed to ensure that low-/mid-frequencies dominate the reconstruction. While unimportant for band-limited imaging data, this could badly affect the performance on LI data, as it prevents us from fully exploiting the unique advantages these short-exposures offer us in constraining the high-resolution estimate of the scene.

To summarise, in practice we find that as long as the blur kernel estimation is accurate, these simple penalisation terms for the scene are either unnecessary, or create more problems than they solve.

\subsubsection{No sum-to-one constraint}

There is a remaining multiplicative degeneracy between the $k_n$ blur kernels and $s$. We have found that successive updates to $s$ will increase its photometric scale, and the scale of the blur kernels will decrease accordingly in order to fit the data (Equation \ref{eq:modelimage}). This does not cause any problems for the results presented here, but for applications where preserving the photometric scale of the data in the reconstructed image is important e.g. (uncalibrated) source flux measurements, a sum-to-one constraint could be applied to the blur kernel pixels. Also, while not being a problem we have encountered, it is conceivable that numerical issues could arise as the scale of the kernels gets smaller, further motivating some sort of regularisation or constraint. For these reasons, we will experiment with implementing such a feature in future applications of our algorithm.

\subsection{Algorithm and Implementation details}\label{sec:alg&imp}

We sketch the pseudo-code for \textit{The Thresher} below (Algorithm \ref{thresh}). Inferring the blur kernel and differential background in step (i) for every $y_n$ image is the computational bottleneck of the algorithm. This problem has received renewed attention in astronomy in the context of Difference Image Analysis (DIA) -- also known as image subtraction -- in the advent of the Legacy Survey of Space and Time \citep{ivezic2019lsst}, and a variety of popular algorithms exist \citep[for example]{alard1998method, bramich2008new, zackay2016proper}. However, these approaches all assume a Gaussian noise model for the imaging data -- which is leveraged to solve the problem analytically -- and cannot be generalised to handle the PGN likelihood (Equation \ref{eq:PGN_likelihood}) appropriate for our short exposure EMCCD images. Consequently, we base our kernel solution approach on the \textit{PyTorchDIA} algorithm \citep{hitchcock2021pytorchdia}, which re-casts the DIA problem as an optimisation, making it general to the choice of objective function. As the name suggests, this code is built within the PyTorch machine learning framework \citep{paszke2019pytorch}, and makes use of the powerful tools within -- such as automatic differentiation and highly optimised, GPU-accelerated convolution computations -- to make it performant.

Like \textit{PyTorchDIA}, \textit{The Thresher} also makes use of PyTorch to automatically compute gradients of the objective function with respect to the image model parameters. We stress again that this does not just bring advantages in computational efficiency. Because of these automatic differentiation capabilities, \textit{The Thresher} is entirely general to the choice of the image model \textit{and} noise model. Although we focus on the case of EMCCD images in this paper, \textit{The Thresher} can straightforwardly generalise to work with other high frame-rate detectors with very different noise properties, such as the next generation of sCMOS devices \citep[]{qiu2013evaluation, steele2016experiments, walker2020will}.

We should note that the convolution of $s$ with a (square) kernel is undefined for pixels in $s$ within half a kernel-width from its edges. Therefore, when fitting $k_n$ and $b_n$ in step (i) of the algorithm, these edge pixels in $s$ and $y_n$ do not enter the loss. Then, in order to ensure that the updated $s$ preserves the original dimensions of the $y_n$ images in step (ii), we must pad $s$ when convolving it with $\hat{k}_n$. In this work, we used a naive zero-padding, and this may result in “edge-effects” in the reconstructed image; while the background level of $s$ should be close to zero, it is not exactly zero, and any sources that are situated on the image edges could be strongly affected. This padding is a delicate but necessary operation, since without it $s$ would become smaller and smaller at each iteration every time it was convolved with a new $\hat{k}_n$.

Also, in order to keep the kernel array size reasonable, we align each $y_n$ image with $s$ to the nearest integer pixel at each iteration. This avoids having to interpolate between pixel values, since any sub-pixel precision mis-alignment between $y_n$ and $s$ will be captured by $\hat{k}_n$ \citep[]{bramich2008new}. While the integer shifts between images should almost always be less than half a kernel width in size, and so will not affect the fit in step (i), any shift, regardless of size will affect the update in step (ii), as "missing" pixels around the image edges must be replaced to preserve the dimensions of each $y_n$ image. As with the aforementioned zero-padding, we again choose to substitute these missing pixels with zeros as a crude estimate of the sky level in the short exposure imaging, which will also contribute to any edge-effects in the reconstructed image.

\makeatletter
\def\BState{\State\hskip-\ALG@thistlm}
\makeatother

\begin{algorithm}
\caption{\textit{The Thresher}}\label{thresh}
\begin{algorithmic}[1]
\State \textbf{Input:} Stream of images $y_n$ for $N \geq 2$
\State \textbf{Output:} Reconstructed image $s$
\State (\textit{scene}) Initialise $s_0$ as the sharpest shift-and-added $X\%$ of TLI images and (optionally) subtract some estimate of its sky background
\State (\textit{scene}) Set $\alpha_0$ so that it is proportional to $s_0$
\State (\textit{kernel}) Set size of kernel array and $\phi$
\State \For{$I$ passes over the data}
\State \While{another image $y_n$ available}
\State Align $y_n$ with $s_t$ to the nearest integer pixel
\State (i)\resizebox{0.79\columnwidth}{!}{$[\hat{k}_n,\hat{b}_n] = \underset{k_n \geq 0, b_n}{\argmin}l(y_n ; k_n, b_n, s_t, \boldsymbol{\Omega}) + n_{\textrm{pix}}\phi || k_n ||_1$}
\State (ii) $s_{t+1}\gets  s_t - \frac{\alpha_0}{\sqrt{\hat{\nu}_t} + \epsilon} \hat{\mu}_t$
\EndWhile
\EndFor
\State \textbf{end}
\State \Return last estimate $s_{t}$

\end{algorithmic}
\end{algorithm}

\section{Simulated Image Tests}\label{sec:simulated}

We first test our algorithm on a detailed simulation of a Lucky Imaging data set. This artificial telescope is similar to the DK154, with the same mirror size, but a pixel scale at 0.05 arcseconds per pixel, rather than the actual DK154s LI camera's 0.09 arcsecond per pixel sampling. The DK154 `red' camera has a broadband filter with an effective wavelength of $\sim 800$ nm, and so we adopt this as the wavelength for this monochromatic simulation.

\subsubsection{Simulating Lucky Imaging data}\label{sec:gen_LI_data}

We used the HCIPy package \citep{por2018high} to generate the atmospheric phase screens needed to simulate the complicated LI PSFs. We adopted a single layer model generated by the infinitely long phase screen extrusion method of \citet{assemat2006method}, which provides a memory efficient approach to generating realistic, time-evolving phase screens as would be observed over the course of a long LI run. Observing conditions were chosen to be similar to those under `good' La Silla seeing conditions, at 0.6 arcseconds, corresponding to a Fried parameter, $r_0 = 27$ cm at 800 nm. Archival GSM data was used to set the outer scale of the turbulence to $\mathcal{L}_0 = 24$ m, and the wind-speed to $v = 5.8$ m/s \citep{martin1998first}.

Planar light waves -- consisting of real and imaginary components -- are propagated through this atmospheric model and are incident on a circular telescope aperture with diameter $D = 1.54$ m. The \emph{instantaneous} speckle-pattern observed on the detector at the focal plane of the telescope is then obtained by applying a fourier transform to this wavefront and down-sampling it to the appropriate pixel scale, at 0.05 arcseconds per pixel.

The \emph{coherence time} for speckle-patterns associated with this single layer atmospheric model can be computed as $\tau_e \sim 0.31 D / v = 80$ ms \citep{tubbs2003lucky}. In practice, the Danish 1.54m usually takes 100 ms exposures, and so some degree of time-averaging will occur. To simulate the PSFs associated with each 100 ms exposure, we therefore time-average 100 speckle patterns sampled at 1 ms intervals.

Equipped with a simulation for the PSF, we can now populate a set of images with point sources in the following way.

(i) We first define a $128\times 128$ pixel grid, and choose this to be populated with 10 point sources.

(ii) Fluxes (in units of $e_{\mathrm{Photon}}^{-}$) are randomly drawn from $U \sim (10, 500)$ for every source.

(iii) The sub-pixel $(x, y)$ positions of these 10 sources are randomly populated across the image, while avoiding the regions close to the edges. The brightest source is moved to the exact centre of the image.

(iv) In each image, the associated 0.1 s time-averaged PSF is scaled to the flux from (ii) at the position of (iii) on the grid defined in (i).

(v) A sky level of 0.1 $e_{\mathrm{Photon}}^{-}$ is then added to each image.

(vi) Finally, for every pixel in each noiseless image, ADU values are randomly sampled from the PGN PDF (Equation \ref{eq:PGN_likelihood}), with $f=25$ ($e_{\mathrm{EM}}^{-} / \textrm{ADU}$), $G=300$ ($e_{\mathrm{EM}}^{-} / e_{\mathrm{Photon}}^{-}$), $\sigma_0 = 60$ ($e_{\mathrm{EM}}^{-}$), $c=0$ ($e_{\mathrm{Photon}}^{-}$) and $q=1$. This assumes that all contributions from spurious charges are zero, the quantum efficiency is flawless, and that instrumental effects such as per-frame bias level drift, have been perfectly corrected for.

In this way, we generate a stack of 3000 simulated 100 ms exposures.

\subsection{Tests on Noiseless Images}\label{sec:noiseless_tests}

Before deploying our algorithm on realistically noisy and coarsely sampled astronomical images, we will first assess what it is capable of learning from idealised, noiseless and well-sampled, high frame-rate imaging data. For this test, we use our spool of 3000 noiseless short exposure PSFs, which are generated at a pixel scale of $\sim$0.013 arcseconds.

\begin{figure*}
    \centering
    \includegraphics[width=\linewidth]{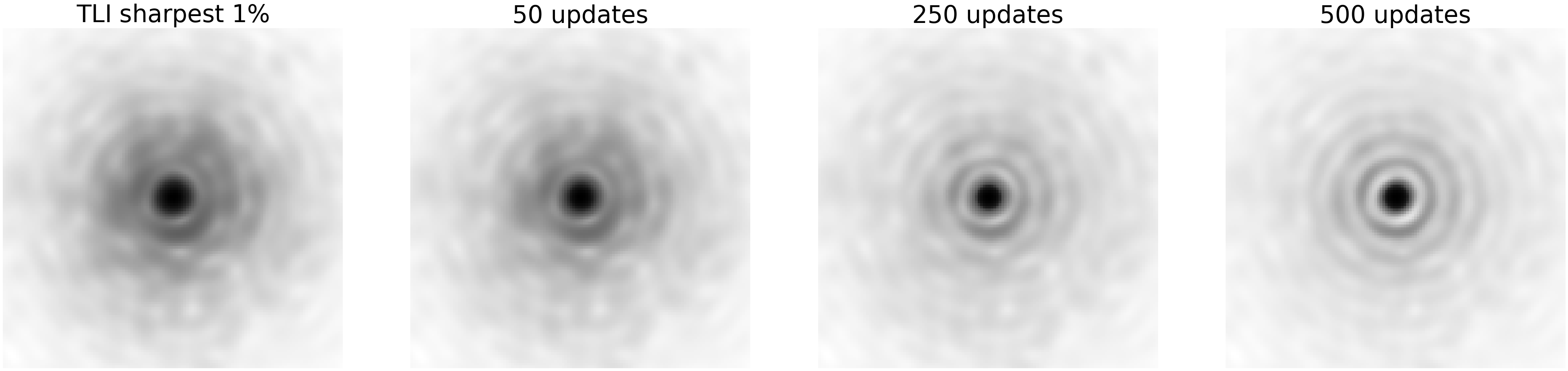}
    \caption{The model image returned by our algorithm after (left to right) 0, 50, 250 and 500 SGD updates from the test on noiseless short exposure PSF images.}
    \label{fig:NoiselessModelUpdates}
\end{figure*}

\begin{figure*}
    \centering
    \includegraphics[width=\linewidth]{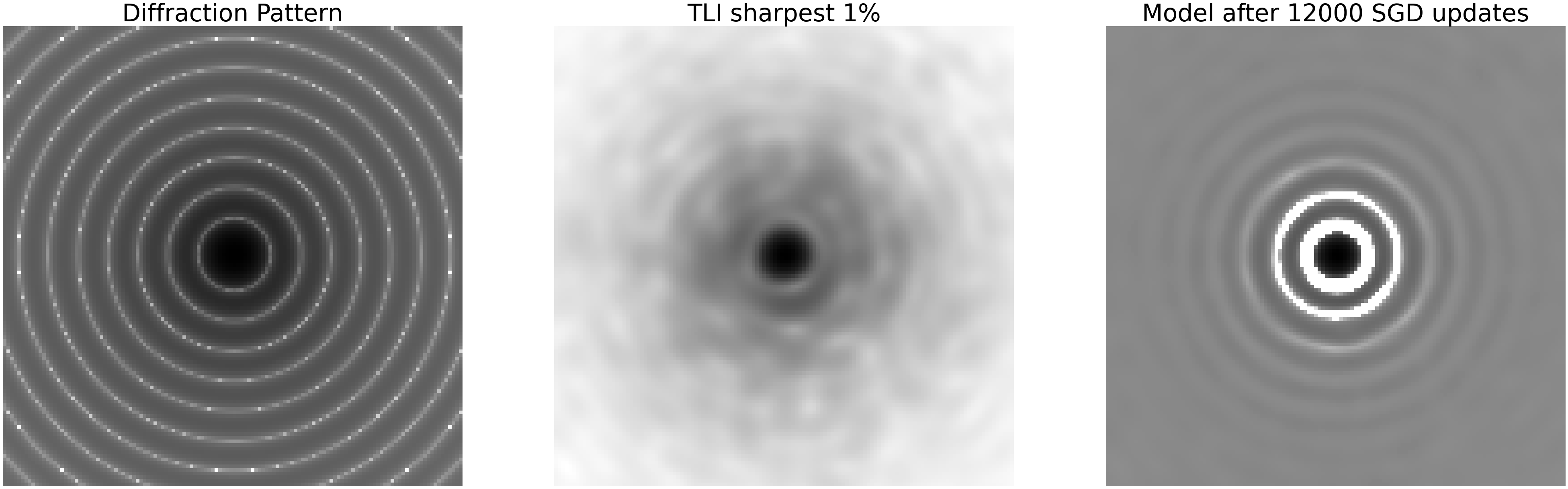}
    \caption{$128\times128$ pixel cutouts of the (from left to right) theoretical instrument diffraction pattern, the shift-and-added sharpest 1\% of images and the model returned by our algorithm after 12000 updates from the test on the 3000 noiseless short exposure PSF images. All cutouts are normalised relative to the highest intensity pixel, and are plotted on a logarithmic scale. The squared error loss used by our algorithm for these noiseless data does not require non-negativity of the model, and so we do not enforce this constraint. Consequently, some regions close to the centre went negative, and for the purposes of plotting on the log-scale we added a small offset.}
    \label{fig:NoiselessTest}
\end{figure*}

The model image was initialised with the sharpest 1\% of shift-and-added exposures. Figure \ref{fig:NoiselessModelUpdates} shows how this estimate of the high-resolution scene improves with successive SGD updates. We ran our algorithm for a total of four passes (equivalent to 12,000 SGD updates) over the noiseless image spool, and in Figure \ref{fig:NoiselessTest}, we plot the diffraction limited Airy pattern, the sharpest 1\% of shift-and-added exposures, and the model returned by our algorithm after 12,000 updates. As the data are noiseless, we adopted a simple squared error for the loss function (i.e. (data - model)$^{2}$). Under this loss function, we do not adopt any non-negativity constraint on the scene model. As the data are noiseless, we do not need to enforce any L1 regularisation on the kernel pixels. The SGD step-size was initialised as $\alpha_0 = 0.005 s_0$, and the kernel array was set to be $129\times129$ pixels large\footnote{In this case of noiseless data and a simple squared error loss function, we found that fitting the large kernel array in step (i) of our algorithm was more quickly achieved with the L-BFGS method \citep{liu1989limited}, and so we used that optimisation algorithm to minimise Equation \ref{eq:step1} in this test, and this test only.} .

\begin{figure*}
    \centering
    \includegraphics[width=\linewidth]{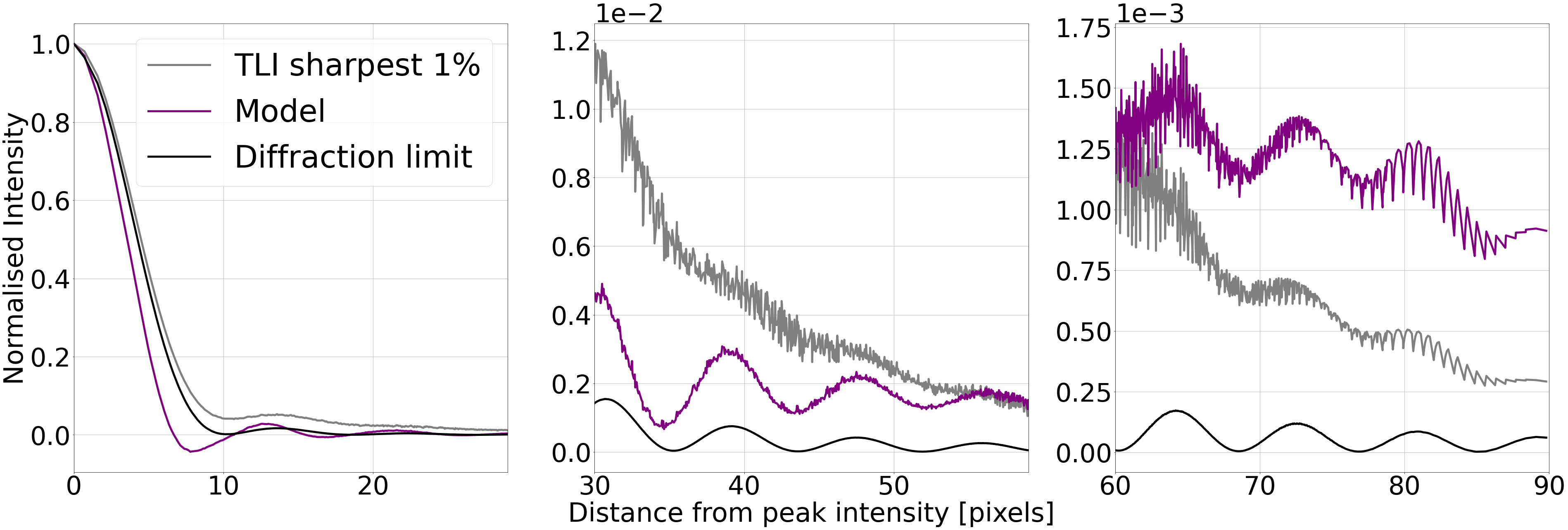}
    \caption{Radial profiles of the PSF images in Figure \ref{fig:NoiselessTest}. Each (normalised) intensity value is the mean of all pixels at the given distance from the peak. The pixel scale is $\sim 0.013$ arcseconds per pixel.}
    \label{fig:NoiselessProfiles}
\end{figure*}

Encouragingly, albeit in this highly idealised scenario, the model returned by \textit{The Thresher} is able to recover something like the Airy diffraction pattern of the telescope. In fact, the central lobe of the model PSF is tighter than that of the instrument diffraction pattern. This can be more clearly seen by plotting the radial profiles of these images in Figure \ref{fig:NoiselessProfiles}. We can see that the peaks of the rings returned by the model are at approximately the same location as those of the diffraction limited PSF, and the intensities of the peaks decreases at about the same rate with distance from the centre. However, for rings closer to the peak intensity, they tend to be shifted slightly towards the centre, and the central lobe itself is noticeably sharper.

It is reasonable to wonder if it really is possible for \textit{The Thresher} to do better than the optics alone. In theory, the information to achieve this does exist in our \emph{collection} of images. Due to atmospheric turbulence, the short exposure PSFs have wandering centroids, and so we get a slightly different view of the scene in each observation; there is additional spatiotemporal data available in the sequence of images. It is this extra information which allows super-resolution methods to reconstruct images sharper than the diffraction limit \citep{borman1998super}.

\subsection{Tests on Noisy Images}\label{sec:noisy_tests}

Next, we run our algorithm on our simulated DK154 EMCCD data (see Section \ref{sec:gen_LI_data} for a description of the how the images are generated). The initialisation for the model, $s_0$, was constructed as the TLI sharpest 50\% of images. The median pixel value of $s_0$ was subtracted as an estimate of the background, and non-negativity was enforced. We set the kernel array size to be $25\times25$ pixels, $\phi = 0.07$, $\alpha_0 = 0.005 s_0$, and we made a single pass over the data set (i.e. 3000 SGD updates).

\begin{figure*}
    \centering
    \includegraphics[width=\linewidth]{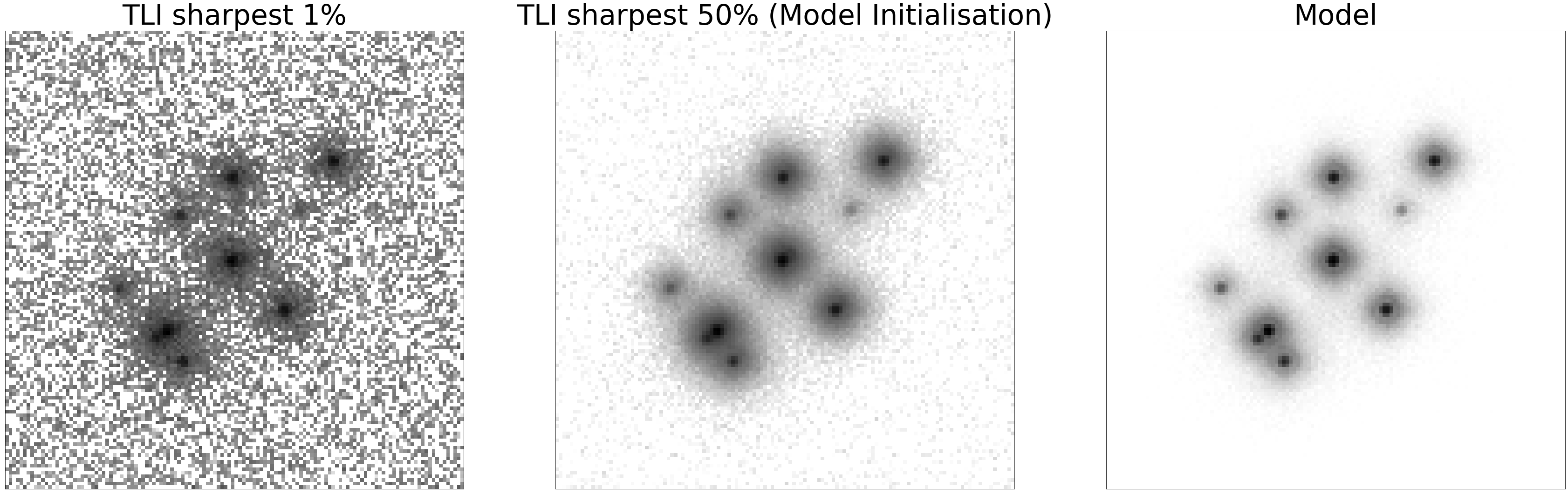}
    \caption{A comparison of the TLI sharpest 1\% and sharpest 50\% of images, and the image model returned by \textit{The Thresher} after a single pass over a stack of 3000 simulated short exposures. For comparison with the model, the TLI coadds have had their respective median pixel values subtracted (as estimates of their sky levels) and non-negativity enforced. All images are on a log-scale.}
    \label{fig:SynTests_ImageComparison}
\end{figure*}

We plot the model image and the TLI sharpest 1\% and 50\% coadds in Figure \ref{fig:SynTests_ImageComparison}; to highlight the noise in the images, they are all displayed on a log-scale, with a small positive offset added for visual clarity. The coadd of the sharpest 1\% images is sharp, but at very low signal-to-noise. The point sources in the coadd of the sharpest 50\% have high signal-to-noise, but this comes at the expense of resolution, and the PSF exhibits a broad, extended halo characteristic of shift-and-add methods.

\begin{figure*}
    \centering
    \includegraphics[width=\linewidth]{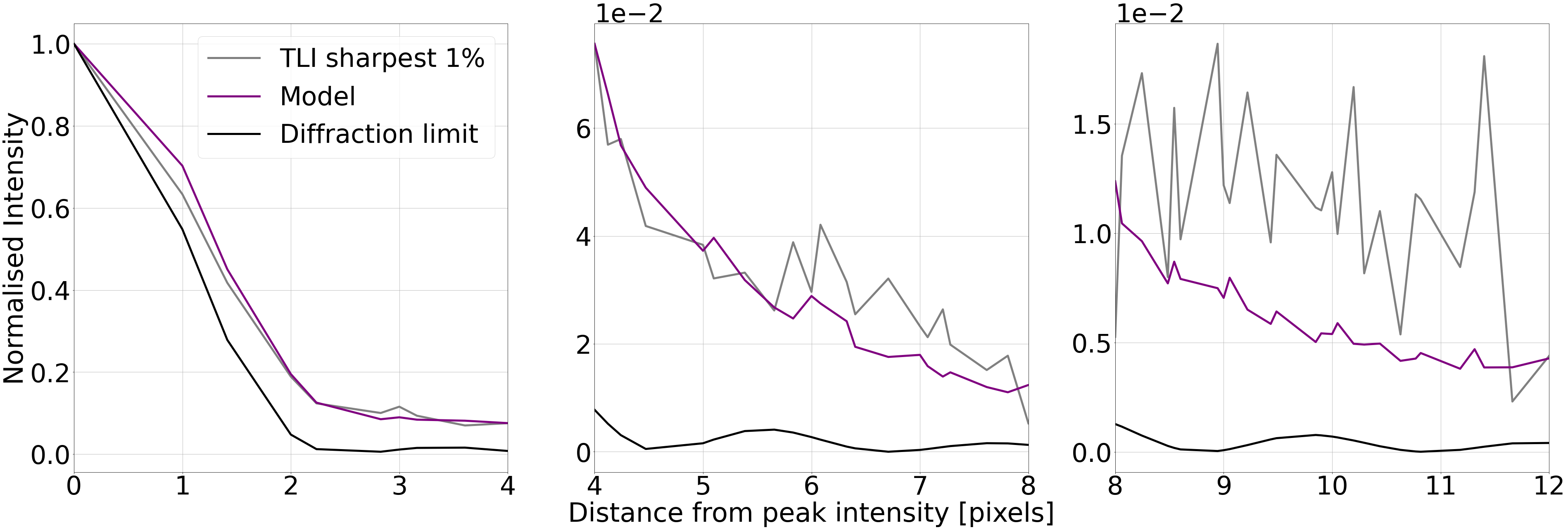}
    \caption{Radial profiles of the bright, central star in Figure \ref{fig:SynTests_ImageComparison} in the sharpest 1\% TLI coadd and the image model returned by \textit{The Thresher}. The theoretical diffraction limited PSF is plotted for comparison. Each (normalised) intensity value is the mean of all pixels at the given distance from the peak. The pixel scale is $\sim 0.05$ arcseconds per pixel.}
    \label{fig:SimulatedProfiles}
\end{figure*}

\textit{The Thresher}, after having been fed 100\% of the images in the spool, delivers the best of both worlds; the image model is as sharp as the best images in the data set, but at substantially higher signal-to-noise. Plots of the radial profile of the central brightest star -- normalised to the same scale -- are shown in Figure \ref{fig:SimulatedProfiles}. The width of the PSF in our image model is comparable in sharpness to the PSF in the TLI sharpest 1\% coadd, but appears much smoother due to the significant improvement in signal-to-noise.

It is important to note that we have not run our algorithm to some converging resolution. Unlike in the highly idealised scenario in Section \ref{sec:noiseless_tests}, as the image model starts to approach something close to the diffraction limited scene, we encounter issues with representing $s$ on this now coarsely sampled grid. This in turn badly affects the per-image kernel inference, leading to a positive feedback loop of bad updates to the image model. This issue could be mitigated by representing $s$ and each $k_n$ at a finer pixel sampling, and including a down-sampling operation in the forward model to match the resolution of the data (i.e. a super-resolution approach). However, for typical astronomical images, the associated computational expense can become prohibitive, and in practice, it is not always necessary to achieve this resolution. The utility of our algorithm is that it can do \textit{better} than TLI in terms of resolution and signal-to-noise. For most science goals, this is generally a more useful objective to aim for than a perfect deconvolution.

\section{Real Image Tests}\label{sec:real}

Here, we run \textit{The Thresher} over a spool of 4800 real 0.1 second DK154 `red' (i.e. $\sim 800$nm wavelength) EMCCD exposures, streaming through the $y_n$ images chronologically. The target is a central $20\times 20$ arcsecond (i.e. $256\times 256$ pixel) region of the globular cluster NGC~7089, and the time-averaged seeing was $\sim 0.8$ arcseconds. We adopt the noise model parameters estimated from the fits to dark images in Section \ref{sec:calibration} to calibrate our likelihood function (Equation \ref{eq:PGN_likelihood}). All raw images were dithered (as described in Section \ref{sec:noisemodelval}), and then bias subtracted (including the per-frame drift level) and flat-fielded. A cosmic ray was identified on a single image in the spool; the affected pixels were fairly isolated from any sources, and so their values were simply replaced with the median value of the image, as an estimate for the sky background.

$s_0$ was again initialised as the shift-and-added sharpest 50\% of images. Next, we subtracted an estimate of the 2D sky background\footnote{This was estimated using the background tools in the photutils package \citep{bradley2016photutils}.} of $s_0$ and enforced non-negativity. The step-size was initialised as $\alpha_0 = 0.005 s_0$, and the tuning constant controlling the L1 regularisation on the kernel was set to $\phi = 0.03$. The kernel array was set to be $25\times 25$ pixels, and we make a single pass over the data (i.e. 4800 SGD updates).

\begin{figure*}
    \centering
    \includegraphics[width=\linewidth]{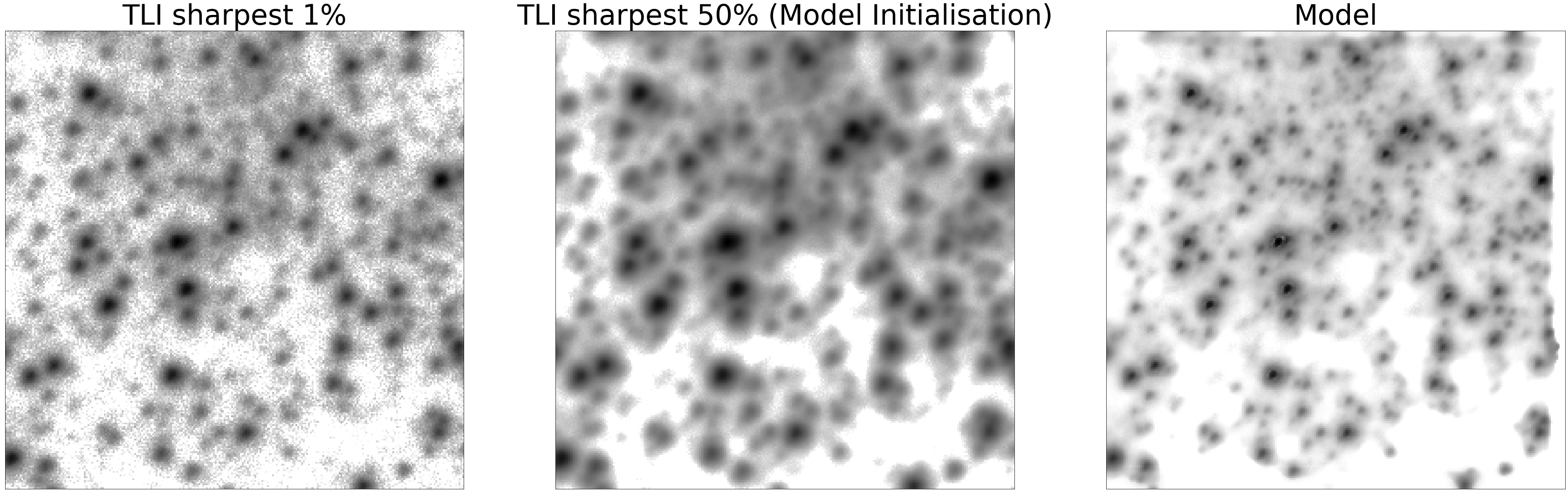}
    \caption{A comparison of the TLI sharpest 1\% and 50\% of images, and the image model returned by \textit{The Thresher} after a single pass over a stack of 4800 real short exposures. For comparison with the model, the TLI coadds have had their respective sky background's subtracted and non-negativity enforced. All images are on a log-scale.}
    \label{fig:RealTests_ImageComparison}
\end{figure*}

\begin{figure*}
    \centering
    \includegraphics[width=\linewidth]{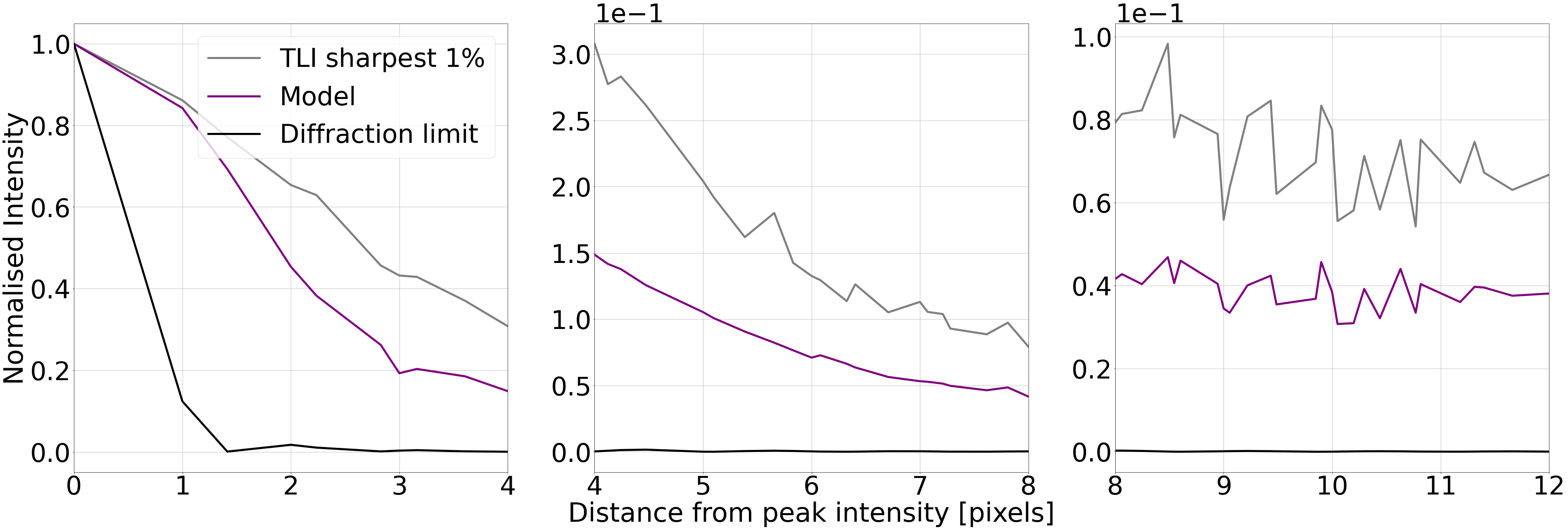}
    \caption{Radial profiles of a bright, fairly isolated, centrally located star in Figure \ref{fig:RealTests_ImageComparison} in the sharpest 1\% TLI coadd and the image model returned by \textit{The Thresher}. The theoretical diffraction limited PSF is plotted for comparison. Each (normalised) intensity value is the mean of all pixels at the given distance from the peak. The pixel scale is $\sim 0.09$ arcseconds per pixel.}
    \label{fig:RealProfiles}
\end{figure*}

In TLI, the SNR of the coadd can only be improved by stacking increasingly blurrier images, and so the resolution rapidly worsens. Provided \textit{The Thresher} is able to accurately account for the individual frame PSFs, feeding more data to the algorithm will not degrade the resolution of the image model and the SNR should increase. This is exactly what we see in Figures \ref{fig:RealTests_ImageComparison} and \ref{fig:RealProfiles}\footnote{In reality, the Danish 1.54m is not diffraction limited, and the limiting resolution is set by the mirror support and optical system. Consequently, the sharpest achievable PSFs in any given short exposures appear somewhat triangular, and the diffraction limit PSF in Figure \ref{fig:RealProfiles} is purely instructive.}; the image model returned by our algorithm is of comparable sharpness to the sharpest 1\% of images in the stack, but at substanstially higher signal-to-noise.

Indeed, many faint sources that are either blended and/or barely distinguishable above the background in the coadds are clearly visible in the image model. This capability of \textit{The Thresher} to reconstruct even very faint sources was an important goal during its design, and this directly motivated the development of the robust SGD procedure and accurate, physically motivated EMCCD noise model.

We note that some sources close to the edges of the TLI coadds cannot be clearly viewed in the model, and there is a prominent absence of flux along the model's right-hand edge. This is due to issues with the convolution operation being undefined for these border pixels, and data images being aligned with the model, as explained in Section \ref{sec:alg&imp}.

\section{Current limitations of \textit{The Thresher} and the scope for future work}\label{sec:limits}

\subsection{Flux Non-linearity}

One attractive property of coadding astronomical images is that source fluxes are preserved. In the general case, where the sky background level of the data are non-zero (i.e. every $b_n > 0$), deconvolution algorithms provide no such guarantee. Historically, this was an area of interest during the early years of the Hubble Space Telescope (HST), where a variety of deconvolution algorithms were used to partially compensate for the aberrations introduced by the defective primary mirror (see \citealt{white1991restoration} for an overview). Some of these studies indicated that the deconvolution introduced systematic non-linearities in flux in the reconstructed image, and so an additional correction was needed to calibrate the photometry. This issue motivated the development of deconvolution algorithms adopting a "two-channel" approach, which have been shown to preserve source fluxes by decomposing the deconvolved image model into point sources and background \citep[for example]{hook1994image, magain1998deconvolution}.

\begin{figure*}
    \centering
    \begin{minipage}{0.5\textwidth}
        \centering
        \includegraphics[width=0.90\textwidth]{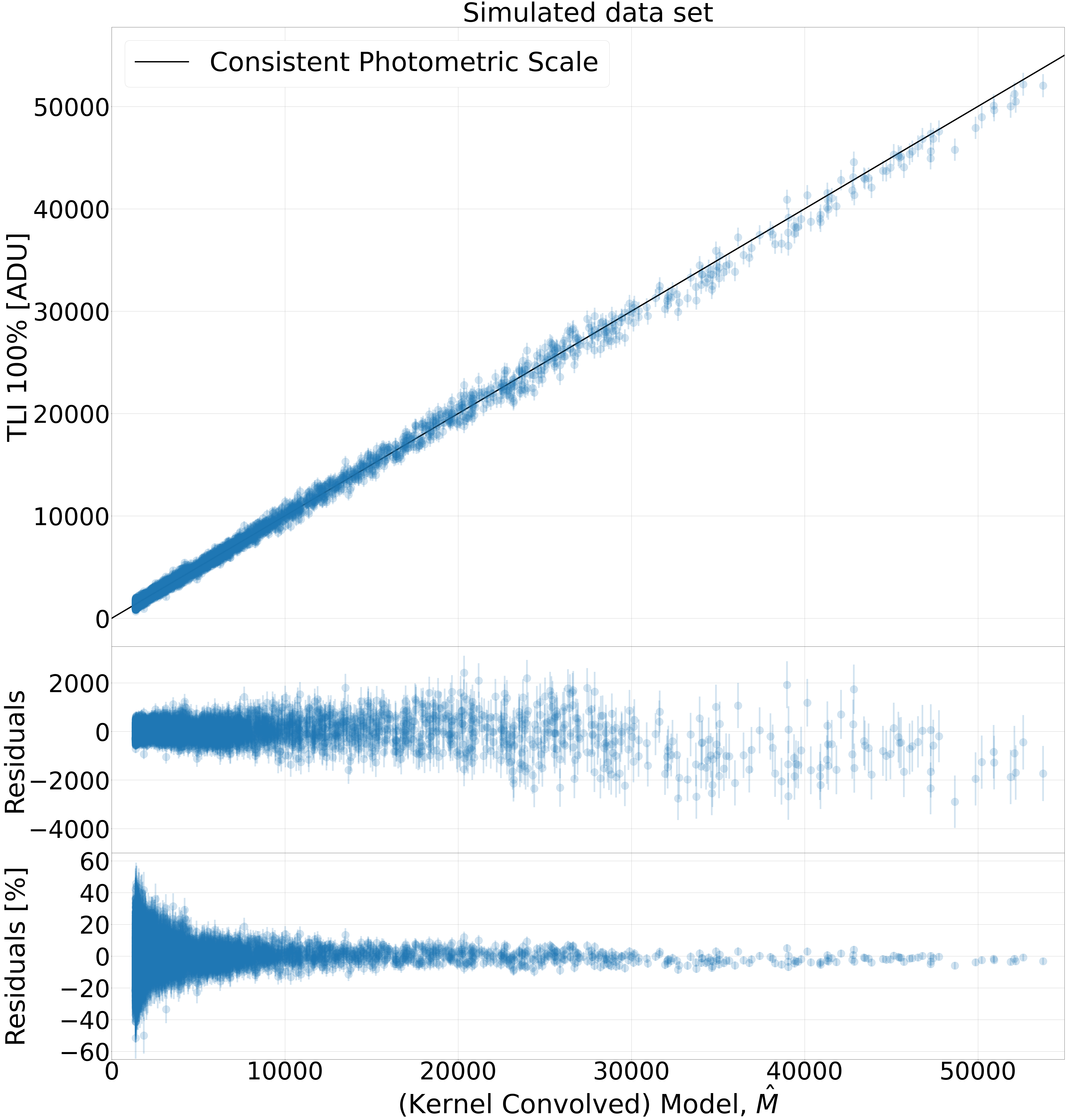}
    \end{minipage}\hfill
    \begin{minipage}{0.5\textwidth}
        \centering
        \includegraphics[width=0.96\textwidth]{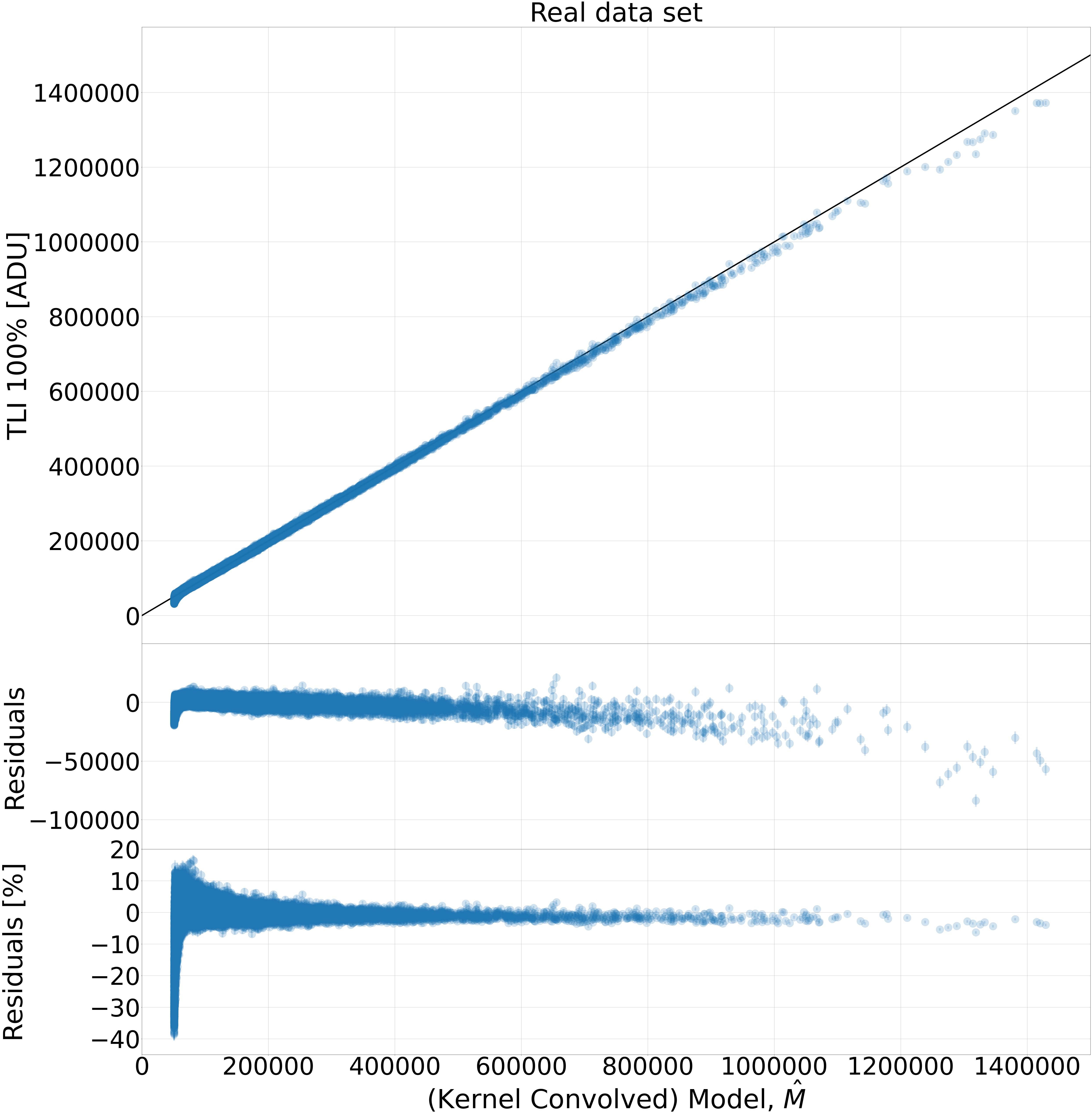}
    \end{minipage}
    \caption{A plot of pixel values of the (kernel convolved) reconstructed image vs. the respective counts in a TLI 100\% selection coadd from the tests in (left) Section \ref{sec:noisy_tests} and  (right) Section \ref{sec:real}. The straight line -- with gradient 1 and intercept 0 -- represents photometric consistency between the model and data as a function of flux. The residuals in the bottom panel are expressed as a percentage of the model value.}
    \label{fig:phot_const}
\end{figure*}

\textit{The Thresher}, as presented in this paper, is not designed to explicitly distinguish between point sources and sky background in the scene, and so we should not expect the relative photometric scale of the data to be preserved in the reconstructed image if the background is non-zero. We probe this behaviour using the results of Sections \ref{sec:noisy_tests} and \ref{sec:real}, where $b_n > 0$ for every $y_n$ image. We do this by fitting image models (of the form in Equation \ref{eq:modelimage}) to the TLI coadds of all the images in each data set. If a reconstructed image has preserved the relative photometric scale of sources in the data, then a plot of pixel values in (the kernel convolved) $s$ against their counterparts in the coadd should be consistent with a straight line with a slope of 1, and intercept 0. As we are now fitting an image model to the shift-and-added stack of the short exposure images, $Y$, we can adopt a Gaussian approximation for its noise, where only the photon contributions are significant. We adopt upper-case notation here to make clear that we are referring to coadds of the $y_n$ short exposures, so Equation \ref{eq:modelimage} now becomes
\begin{equation}
    M_{ij} = [K \otimes s]_{ij} + B \;.
    \label{eq:modelimage_coadd}
\end{equation}
Ignoring the irrelevant normalisation constant, the negative log-likelihood takes the form,
\begin{multline}
    l_{\textrm{N}} \left( \, Y \, ; \, K, B, s, f, G \, \right)
    = - \sum_{ij} \ln p_{\textrm{N}}\left( \, Y_{ij} \, | \, K, B, s, f, G \, \right) \\
    = \frac{1}{2}\sum_{ij}\left(\frac{Y_{ij} - M_{ij}}{\sigma_{ij}} \right)^2 \;\ + \sum_{ij} \ln\;\sigma_{ij}  \;,
    \label{eq:normal_likelihood}
\end{multline}
with photon shot-noise limited per-pixel uncertainties\footnote{Note that although we are in the Gaussian limit, there is no analytical approach to minimising this function due to the photon shot noise. The factor of two in Equation \ref{eq:normal_uncertainties} is referred to as the `excess noise' factor, and accounts for the probabilistic cascade amplification \citep[]{korevaar2011maximum, hirsch2013stochastic}.}
\begin{equation}
    \sigma_{ij}^{2} = 2 M_{ij}\frac{G}{f} \;.
    \label{eq:normal_uncertainties}
\end{equation}
We can then return the MLEs of the PSF-matching kernel and sky background that fit $s$ to the coadd,
\begin{equation}
    [\hat{K}, \; \hat{B}] = \argmin_{K, \; B} l_{\textrm{N}}(Y ; K, B, s, f, G)
    \label{eq:DIA}
\end{equation}
and compute $\hat{M}$ via Equation \ref{eq:modelimage_coadd}. We do this for both of the final estimates of $s$ returned by the tests on simulated and real images (Sections \ref{sec:noisy_tests} and \ref{sec:real}), fitting each of these $s$ estimates to their respective stacks of 100\% of the shift-and-added data. Plots of pixel counts in the image model, $\hat{M}$, against their respective counts in the coadds are shown in Figure \ref{fig:phot_const}. In the middle and bottom panels we plot the residuals from the straight line representing photometric consistency between the model and data; residuals in the bottom panels are expressed as a percentage of the model value (i.e. as a percentage of the model brightness). Uncertainties on the points are equal to the square root of the variances given by Equation \ref{eq:normal_uncertainties}.

The two plots suggest that \textit{The Thresher} does indeed introduce non-linearities into the flux of the reconstructed image. Specifically, fluxes of bright objects are systematically overestimated; in both plots, the residuals from a consistent photometric scale become more negative with increasing model brightness. This is most clearly seen in the right panel of Figure \ref{fig:phot_const}, as the dynamic range of sources in the crowded field is large. The asymmetry in the residuals at the faint end is an artefact of the 2D background subtraction of the scene initialisation.

Some experimentation has suggested that this problem improves by allowing \textit{The Thresher} to run for longer, making several passes over the data (similar to findings by \citealt{lindler2013interpretation}). However, this has to be balanced with computational expense and issues with model representation as the reconstructed sources become increasingly sharper. Given the importance of preserving the local photometric scale of the data in the reconstructed image for many science cases, this is highlighted as something to address for future work on our algorithm.

\subsection{Computational expense}

The computational bottleneck of our algorithm is set by the speed with which we can fit $k_n$ and $b_n$ (Equation \ref{eq:step1_penalised}), as this requires computing a convolution for the estimate of the forward model (Equation \ref{eq:modelimage}) at each step in the optimisation as $l$ is minimised. The convolution operation is embarrassingly parallel, and so \textit{The Thresher} leverages GPUs if they are available.

Lucky Imaging data sets consist of thousands of images, but since the algorithm only ever has to access a single image at a time, memory management is not an issue. Nonetheless, it must process these many images, and so must perform \emph{very many} convolutions. All tests in this paper were run on modest hardware -- a laptop with a single GeForce GTX 1050 -- and so execution times were fairly slow, at about $2.5$, $6$ and $20$ hours for the tests in Sections \ref{sec:noisy_tests}, \ref{sec:real} and \ref{sec:noiseless_tests} respectively. Desktop GPUs will of course help to beat-down those times, particularly if the images are large, but the technique remains computationally expensive. Indeed, although the code is designed to be agnostic to the hardware, if only a CPU is available to the user, it may prove to be computationally intractable for most problems.

\section{Conclusions}\label{sec:conclusion}

We have presented \textit{The Thresher}, a new algorithm for processing Lucky Imaging data. It fundamentally differs from Traditional Lucky Imaging (TLI) shift-and-add procedures in that it optimises a physically motivated likelihood function for the entire set of imaging data to recover the underlying astronomical scene. Because it uses the full data set, \textit{The Thresher} outperforms TLI in signal-to-noise; because it accounts for the individual-frame PSFs, it outperforms TLI in angular resolution.

We implement an accurate noise model for low light level, Electron Multiplying CCD data. When combined with a robust stochastic gradient descent procedure and appropriate regularisation of model parameters, \textit{The Thresher} can cleanly reconstruct even extremely faint sources, which may be below the detection threshold in any given short exposure. \textit{The Thresher} is entirely general to the choice of image model and noise model, as it makes use of automatic differentiation tools, and has been designed to anticipate near future high-frame rate imaging on sCMOS devices. Furthermore, it makes use of GPUs to massively accelerate the many convolution computations.

While the image model returned by the current version of \textit{The Thresher} can be used for, among other things, astrometry and source detection, systematic non-linearities in flux must be corrected for if accurate photometry is needed. Addressing this issue will be a focus for future development of our algorithm.

We end by commenting that the ideas in \textit{The Thresher}, while particularly well suited to high frame-rate imaging data, can also be useful for fitting models to conventional images. To this end, we also include the option to adopt a Gaussian log-likelihood for the noise in the imaging (of the form in Equation \ref{eq:normal_likelihood}) suitable for conventional, ground-based CCD exposures, as part of our software implementation. While the deconvolution would no longer benefit from the high-frequency information uniquely available to short exposures, \textit{The Thresher} provides an interesting alternative to co-addition approaches when the image-to-image PSF is varying.

Development of the software implementation of our algorithm is ongoing, and we refer the reader to the following Github repository \url{https://github.com/jah1994/TheThresher}.

\section*{Acknowledgements}

We would like to thank the MiNDSTEp consortium, and Jesper Skottfelt in particular, for providing the Danish 1.54m data used to both calibrate the detector noise model and test our algorithm. J. A. Hitchcock acknowledges funding
from the Science and Technology Facilities Council of the United
Kingdom.

\section*{Data Availability}

The data underlying this article will be shared on reasonable request
to the corresponding author.



\bibliographystyle{mnras}
\bibliography{Thresher} 

\begin{thebibliography}{}
\makeatletter
\relax
\def\mn@urlcharsother{\let\do\@makeother \do\$\do\&\do\#\do\^\do\_\do\%\do\~}
\def\mn@doi{\begingroup\mn@urlcharsother \@ifnextchar [ {\mn@doi@}
  {\mn@doi@[]}}
\def\mn@doi@[#1]#2{\def\@tempa{#1}\ifx\@tempa\@empty \href
  {http://dx.doi.org/#2} {doi:#2}\else \href {http://dx.doi.org/#2} {#1}\fi
  \endgroup}
\def\mn@eprint#1#2{\mn@eprint@#1:#2::\@nil}
\def\mn@eprint@arXiv#1{\href {http://arxiv.org/abs/#1} {{\tt arXiv:#1}}}
\def\mn@eprint@dblp#1{\href {http://dblp.uni-trier.de/rec/bibtex/#1.xml}
  {dblp:#1}}
\def\mn@eprint@#1:#2:#3:#4\@nil{\def\@tempa {#1}\def\@tempb {#2}\def\@tempc
  {#3}\ifx \@tempc \@empty \let \@tempc \@tempb \let \@tempb \@tempa \fi \ifx
  \@tempb \@empty \def\@tempb {arXiv}\fi \@ifundefined
  {mn@eprint@\@tempb}{\@tempb:\@tempc}{\expandafter \expandafter \csname
  mn@eprint@\@tempb\endcsname \expandafter{\@tempc}}}

\bibitem[\protect\citeauthoryear{Alard \& Lupton}{Alard \&
  Lupton}{1998}]{alard1998method}
Alard C.,  Lupton R.~H.,  1998, The Astrophysical Journal, 503, 325

\bibitem[\protect\citeauthoryear{Ass{\'e}mat, Wilson  \& Gendron}{Ass{\'e}mat
  et~al.}{2006}]{assemat2006method}
Ass{\'e}mat F.,  Wilson R.~W.,   Gendron E.,  2006, Optics express, 14, 988

\bibitem[\protect\citeauthoryear{Ayers \& Dainty}{Ayers \&
  Dainty}{1988}]{ayers1988iterative}
Ayers G.,  Dainty J.~C.,  1988, Optics letters, 13, 547

\bibitem[\protect\citeauthoryear{Becker, Homrighausen, Connolly, Genovese,
  Owen, Bickerton  \& Lupton}{Becker et~al.}{2012}]{becker2012regularization}
Becker A.,  Homrighausen D.,  Connolly A.,  Genovese C.,  Owen R.,  Bickerton
  S.,   Lupton R.,  2012, Monthly Notices of the Royal Astronomical Society,
  425, 1341

\bibitem[\protect\citeauthoryear{Borman \& Stevenson}{Borman \&
  Stevenson}{1998}]{borman1998super}
Borman S.,  Stevenson R.~L.,  1998, in 1998 Midwest symposium on circuits and
  systems (Cat. No. 98CB36268). pp 374--378

\bibitem[\protect\citeauthoryear{Bottou et~al.}{Bottou
  et~al.}{1998}]{bottou1998online}
Bottou L.,  et~al., 1998, On-line learning in neural networks, 17, 142

\bibitem[\protect\citeauthoryear{Bottou, Curtis  \& Nocedal}{Bottou
  et~al.}{2018}]{bottou2018optimization}
Bottou L.,  Curtis F.~E.,   Nocedal J.,  2018, Siam Review, 60, 223

\bibitem[\protect\citeauthoryear{Bradley et~al.,}{Bradley
  et~al.}{2016}]{bradley2016photutils}
Bradley L.,  et~al., 2016, Astrophysics Source Code Library, pp ascl--1609

\bibitem[\protect\citeauthoryear{Bramich}{Bramich}{2008}]{bramich2008new}
Bramich D.,  2008, Monthly Notices of the Royal Astronomical Society: Letters,
  386, L77

\bibitem[\protect\citeauthoryear{Bramich, Horne, Alsubai, Bachelet, Mislis  \&
  Parley}{Bramich et~al.}{2016}]{bramich2016difference}
Bramich D.,  Horne K.,  Alsubai K.,  Bachelet E.,  Mislis D.,   Parley N.,
  2016, Monthly Notices of the Royal Astronomical Society, 457, 542

\bibitem[\protect\citeauthoryear{Campisi \& Egiazarian}{Campisi \&
  Egiazarian}{2017}]{campisi2017blind}
Campisi P.,  Egiazarian K.,  2017, Blind image deconvolution: theory and
  applications.
CRC press

\bibitem[\protect\citeauthoryear{Harps{\o}e, J{\o}rgensen, Andersen  \&
  Grundahl}{Harps{\o}e et~al.}{2012}]{harpsoe2012high}
Harps{\o}e K.~B.,  J{\o}rgensen U.~G.,  Andersen M.~I.,   Grundahl F.,  2012,
  Astronomy \& Astrophysics, 542, A23

\bibitem[\protect\citeauthoryear{Hirsch, Harmeling, Sra  \&
  Sch{\"o}lkopf}{Hirsch et~al.}{2011}]{hirsch2011online}
Hirsch M.,  Harmeling S.,  Sra S.,   Sch{\"o}lkopf B.,  2011, Astronomy \&
  Astrophysics, 531, A9

\bibitem[\protect\citeauthoryear{Hirsch, Wareham, Martin-Fernandez, Hobson  \&
  Rolfe}{Hirsch et~al.}{2013}]{hirsch2013stochastic}
Hirsch M.,  Wareham R.~J.,  Martin-Fernandez M.~L.,  Hobson M.~P.,   Rolfe
  D.~J.,  2013, PloS one, 8, e53671

\bibitem[\protect\citeauthoryear{Hitchcock, Hundertmark, Foreman-Mackey,
  Bachelet, Dominik, Street  \& Tsapras}{Hitchcock
  et~al.}{2021}]{hitchcock2021pytorchdia}
Hitchcock J.~A.,  Hundertmark M.,  Foreman-Mackey D.,  Bachelet E.,  Dominik
  M.,  Street R.,   Tsapras Y.,  2021, Monthly Notices of the Royal
  Astronomical Society, 504, 3561

\bibitem[\protect\citeauthoryear{Hook \& Lucy}{Hook \&
  Lucy}{1994}]{hook1994image}
Hook R.,  Lucy L.,  1994, in The Restoration of HST Images and Spectra-II.
  p.~86

\bibitem[\protect\citeauthoryear{Ivezi{\'c} et~al.,}{Ivezi{\'c}
  et~al.}{2019}]{ivezic2019lsst}
Ivezi{\'c} {\v{Z}}.,  et~al., 2019, The Astrophysical Journal, 873, 111

\bibitem[\protect\citeauthoryear{Kingma \& Ba}{Kingma \&
  Ba}{2014}]{kingma2014adam}
Kingma D.~P.,  Ba J.,  2014, arXiv preprint arXiv:1412.6980

\bibitem[\protect\citeauthoryear{Knuth}{Knuth}{2006}]{knuth2006optimal}
Knuth K.~H.,  2006, arXiv preprint physics/0605197

\bibitem[\protect\citeauthoryear{Korevaar, Goorden, Heemskerk  \&
  Beekman}{Korevaar et~al.}{2011}]{korevaar2011maximum}
Korevaar M.~A.,  Goorden M.~C.,  Heemskerk J.~W.,   Beekman F.~J.,  2011,
  Physics in Medicine \& Biology, 56, 4785

\bibitem[\protect\citeauthoryear{Labeyrie}{Labeyrie}{1970}]{labeyrie1970attainment}
Labeyrie A.,  1970, Astron. Astrophys., 6, 85

\bibitem[\protect\citeauthoryear{Law, Mackay  \& Baldwin}{Law
  et~al.}{2006}]{law2006lucky}
Law N.~M.,  Mackay C.~D.,   Baldwin J.~E.,  2006, Astronomy \& Astrophysics,
  446, 739

\bibitem[\protect\citeauthoryear{Lee, Budav{\'a}ri, White  \& Gulian}{Lee
  et~al.}{2017}]{lee2017robust}
Lee M.~A.,  Budav{\'a}ri T.,  White R.~L.,   Gulian C.,  2017, Astronomy and
  computing, 21, 15

\bibitem[\protect\citeauthoryear{Levin, Weiss, Durand  \& Freeman}{Levin
  et~al.}{2011}]{levin2011understanding}
Levin A.,  Weiss Y.,  Durand F.,   Freeman W.~T.,  2011, IEEE transactions on
  pattern analysis and machine intelligence, 33, 2354

\bibitem[\protect\citeauthoryear{Lindler, A’Hearn, Besse, Carcich, Hermalyn
  \& Klaasen}{Lindler et~al.}{2013}]{lindler2013interpretation}
Lindler D.~J.,  A’Hearn M.~F.,  Besse S.,  Carcich B.,  Hermalyn B.,
  Klaasen K.~P.,  2013, Icarus, 222, 571

\bibitem[\protect\citeauthoryear{Liu \& Nocedal}{Liu \&
  Nocedal}{1989}]{liu1989limited}
Liu D.~C.,  Nocedal J.,  1989, Mathematical programming, 45, 503

\bibitem[\protect\citeauthoryear{Mackay}{Mackay}{2013}]{mackay2013high}
Mackay C.,  2013, Monthly Notices of the Royal Astronomical Society, 432, 702

\bibitem[\protect\citeauthoryear{Magain, Courbin  \& Sohy}{Magain
  et~al.}{1998}]{magain1998deconvolution}
Magain P.,  Courbin F.,   Sohy S.,  1998, The Astrophysical Journal, 494, 472

\bibitem[\protect\citeauthoryear{Martin, Tokovinin, Ziad, Conan, Borgnino,
  Avila, Agabi  \& Sarazin}{Martin et~al.}{1998}]{martin1998first}
Martin F.,  Tokovinin A.,  Ziad A.,  Conan R.,  Borgnino J.,  Avila R.,  Agabi
  A.,   Sarazin M.,  1998, Astronomy and Astrophysics, 336, L49

\bibitem[\protect\citeauthoryear{Paszke et~al.,}{Paszke
  et~al.}{2019}]{paszke2019pytorch}
Paszke A.,  et~al., 2019, arXiv preprint arXiv:1912.01703

\bibitem[\protect\citeauthoryear{Por, Haffert, Radhakrishnan, Doelman, van
  Kooten  \& Bos}{Por et~al.}{2018}]{por2018high}
Por E.~H.,  Haffert S.~Y.,  Radhakrishnan V.~M.,  Doelman D.~S.,  van Kooten
  M.,   Bos S.~P.,  2018, in Adaptive Optics Systems VI. p. 1070342

\bibitem[\protect\citeauthoryear{Qiu, Mao, Lu, Xiang  \& Jiang}{Qiu
  et~al.}{2013}]{qiu2013evaluation}
Qiu P.,  Mao Y.-N.,  Lu X.-M.,  Xiang E.,   Jiang X.-J.,  2013, Research in
  Astronomy and Astrophysics, 13, 615

\bibitem[\protect\citeauthoryear{Skottfelt et~al.,}{Skottfelt
  et~al.}{2015}]{skottfelt2015two}
Skottfelt J.,  et~al., 2015, Astronomy \& Astrophysics, 574, A54

\bibitem[\protect\citeauthoryear{Staley, Mackay, King, Suess  \& Weller}{Staley
  et~al.}{2010}]{staley2010data}
Staley T.~D.,  Mackay C.~D.,  King D.,  Suess F.,   Weller K.,  2010, in
  Ground-based and Airborne Instrumentation for Astronomy III. p. 77355Z

\bibitem[\protect\citeauthoryear{Steele, Jermak, Copperwheat, Smith,
  Poshyachinda  \& Soonthorntham}{Steele et~al.}{2016}]{steele2016experiments}
Steele I.~A.,  Jermak H.,  Copperwheat C.~M.,  Smith R.~J.,  Poshyachinda S.,
  Soonthorntham B.,  2016, in High Energy, Optical, and Infrared Detectors for
  Astronomy VII. p. 991522

\bibitem[\protect\citeauthoryear{Tubbs}{Tubbs}{2003}]{tubbs2003lucky}
Tubbs R.~N.,  2003, arXiv preprint astro-ph/0311481

\bibitem[\protect\citeauthoryear{Vio, Bardsley  \& Wamsteker}{Vio
  et~al.}{2005}]{vio2005least}
Vio R.,  Bardsley J.,   Wamsteker W.,  2005, Astronomy \& Astrophysics, 436,
  741

\bibitem[\protect\citeauthoryear{Walker}{Walker}{2020}]{walker2020will}
Walker G.,  2020, in American Astronomical Society Meeting Abstracts\# 235. pp
  175--01

\bibitem[\protect\citeauthoryear{White \& Allen}{White \&
  Allen}{1991}]{white1991restoration}
White R.,  Allen R.,  1991, The Restoration of HST Images and Spectra

\bibitem[\protect\citeauthoryear{Zackay, Ofek  \& Gal-Yam}{Zackay
  et~al.}{2016}]{zackay2016proper}
Zackay B.,  Ofek E.~O.,   Gal-Yam A.,  2016, The Astrophysical Journal, 830, 27

\makeatother
\end{thebibliography}



\appendix

\section{Table of symbols}

\begin{table*}
    \centering
    \begin{tabular}{c|c}
    \hline
    \textbf{Symbol} & \textbf{Definition}  \\
    \hline
    \textit{Variables} \\
    $\boldsymbol{\theta}$ & vector of image model parameters \\
    $m$ & model for an image \\
    $k$ & blur kernel \\
    $s$ & scene model \\
    $b$ & sky background \\
    $\boldsymbol{y}$ & vector of images \\
    $y$ & single image \\
    $N$ & total number of images \\
    $n_{\textrm{pix}}$ & total number of pixels in a single image \\
    $L$ & total loss for imaging data \\
    $l$ & loss for a single image \\
    $\phi$ & hyper-parameter for tuning the L1 regularisation on $k$ \\
    $\alpha_0$ & SGD learning rate \\
    $\hat{\mu}$ & exponentially decaying average of the gradients of the loss w.r.t $s$ \\
    $\hat{\nu}$ & exponentially decaying average of the squared gradients of the loss w.r.t $s$ \\
    $\boldsymbol{\Omega}$ & vector of noise model parameters \\
    $f$ & A/D conversion factor \\
    $G$ & electron-multiplying (EM) gain \\
    $\sigma_0$ & readout noise \\
    $c$ & spurious charge \\
    $q$ & quantum efficiency \\
    $\lambda$ & rate parameter of the Poisson distribution \\
    $Y$ & image co-add \\
    $M$ & model for an image (image co-add) \\
    $K$ & blur kernel (image co-add) \\
    $B$ & sky background (image co-add) \\
    $\sigma$ & pixel uncertainties (image co-add) \\
    \hline
    \textit{Operators and special functions} \\
    $\otimes$ & convolution \\
    $\nabla$ & vector differential operator \\
    $|| \cdot ||_1$ & L1 norm \\
    $H$ & Heaviside step function \\
    $I_1$ & modified Bessel function of the first order \\
    \hline
    \textit{Iterables and indices} \\
    $I$ & number of complete passes over all images \\
    $n$ & single image index \\
    $t$ & SGD update index \\
    $i$ & image row pixel index \\
    $j$ & image column pixel index \\
    \hline
    \end{tabular}
    \caption{A table of notation used in this paper.}
    \label{tab:notation}
\end{table*}


\bsp	
\label{lastpage}
\end{document}